\documentclass[twocolumn]{elsart}
\usepackage{latexsym, setspace}
\usepackage{epsfig, times}
\begin{document}
\renewcommand{\arraystretch}{1.2}
\begin{frontmatter}
\title{The Forward Time Projection Chamber (FTPC) in STAR}
\collab{
K.H.~Ackermann$^5$, F.~Bieser$^4$, F.P.~Brady$^2$, D.~Cebra$^2$, J.E.~Draper$^2$,
V.~Eckardt$^5$, T.~Eggert$^5$, H.~Fessler$^5$, K.J.~Foley$^1$, V.~Ghazikhanian$^3$,
T.J.~Hallman$^1$, M.~Heffner$^2$, H.~H\"ummler$^5$, J.~Klay$^2$, S.R.~Klein$^4$, 
A.~Lebedev$^1$, M.J.~LeVine$^1$, T.~Ljubicic$^1$, G.~Lo~Curto$^5$, R.S.~Longacre$^1$,
M.~Oldenburg$^5$, HG.~Ritter$^4$, J.L.~Romero$^2$, N.~Schmitz$^5$, A.~Sch\"uttauf$^5$,
J.~Seyboth$^5$, P.~Seyboth$^5$, M.~Vidal$^5$, C.~Whitten Jr.~$^3$, E.~Yamamoto$^3$
}

\address{
$^1$Brookhaven National Laboratory, Upton, New York 11973, USA \\
$^2$University of California, Davis, California 95616, USA \\
$^3$University of California, Los Angeles, California 90095, USA \\
$^4$Lawrence Berkeley National Laboratory, Berkeley, California 94720, USA \\
$^5$Max-Planck-Institut f\"ur Physik, 80805 M\"unchen, Germany
}

\begin{abstract}

Two cylindrical forward TPC detectors are described which were constructed 
to extend the phase space coverage of the STAR experiment to the
region $2.5 < |\eta |<4.0$. For optimal use of the available space and 
in order to cope with the high track density of central Au+Au collisions
at RHIC, a novel design was developed using radial drift in a low diffusion gas. 
From prototype measurements a 2-track resolution of 1 - 2 mm is expected.

\end{abstract}

\end{frontmatter}

\section{Introduction}

The Forward Time Projection Chambers (FTPC) were constructed to extend the 
acceptance of the STAR experiment \cite{Bie98}. 
They cover the pseudorapidity range of $2.5 < |\eta |<4.0$
on both sides of STAR and measure momenta and production rates of positively
and negatively charged particles as well as 
neutral strange particles. Also, due to the high multiplicity, approximately 1000 charged
particles in a central Au+Au collision, event-by-event observables like 
$\langle p_T \rangle$, fluctuations of charged particle multiplicity and 
collective flow anisotropies can be studied. The increased acceptance improves the
general event characterization in STAR and allows the study of asymmetric systems like p+A
collisions. The design and construction was carried out by the group from MPI Munich with 
contributions from LBNL Berkeley, BNL Brookhaven, UC Davis, UCLA Los Angeles, and
MEPhI Moscow \cite{Bie98,Sch99}. 

\section{Detector Design}

\subsection{Conceptual Design}

\begin{figure*}
\epsfig{file=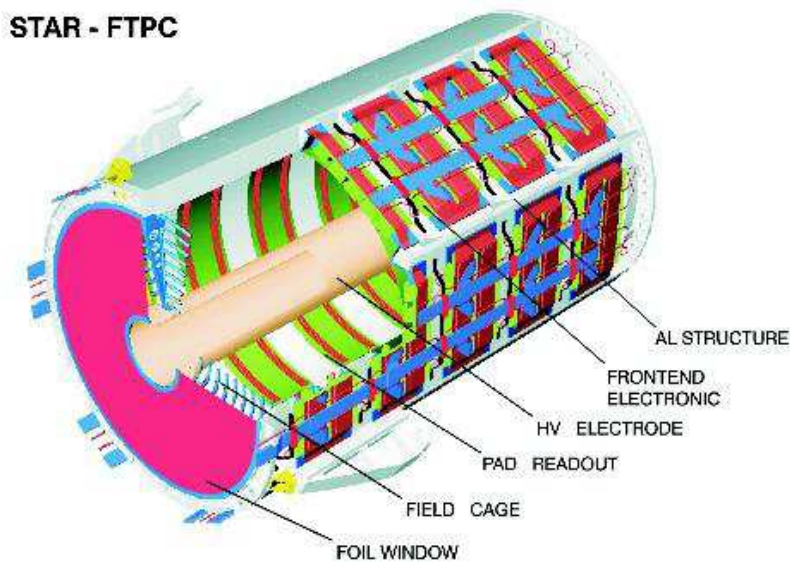,width=\linewidth}
\caption{Schematic diagram of an FTPC for the STAR experiment}
\label{fig1}
\end{figure*}

The FTPC concept was determined mainly by two considerations: Firstly by the high particle
density with tracks under small angles with respect to the beam direction and 
secondly by the restricted available space inside the TPC \cite{Wie97}, where the FTPCs
are located. In Fig. \ref{fig1} the final design is shown. It is a cylindrical structure,
75 cm in diameter and 120 cm long, with a radial drift field and readout chambers
located in 5 rings on the outer cylinder surface. Each ring has two padrows
and is subdivided azimuthally into 6 readout chambers. The radial drift 
configuration was chosen to improve 
the two-track separation in the region close to the beam pipe where the particle
density is highest. The field cage is formed by the inner HV-electrode, 
a thin metalized plastic tube, and the outer cylinder wall at ground potential.
The field region at both ends is closed by a planar structure of concentric rings,
made of thin aluminum pipes. The front end electronics (FEE), which amplifies,
shapes, and digitizes the signals, is mounted on the back of the readout chambers.
Each particle trajectory is sampled up to 10 times. The ionization electrons are
drifted to the anode sense wires and induced signals on the adjacent cathode
surface are read out by 9600 pads (each 1.6$\times$20 mm$^2$).
The above design has some unusual and new features for a TPC: 
\begin{itemize}
  \item The electrons drift
        in a radial electrical field perpendicular to the solenoidal magnetic field.
  \item Curved readout chambers are used to keep the radial field as ideal as possible.
  \item A two-track separation of 1-2 mm is expected, which is an order of
        magnitude better than in all previously built TPCs with pad readout.
\end{itemize} 
To meet these requirements a R+D 
program was initiated, including the selection of the most suitable gas mixture, the
development of the fabrication technology for the curved readout chambers, and the
optimization of the wire and pad geometry for the readout chambers.

\subsection{Selection of Gas Mixture}

\begin{figure*}
\epsfig{file=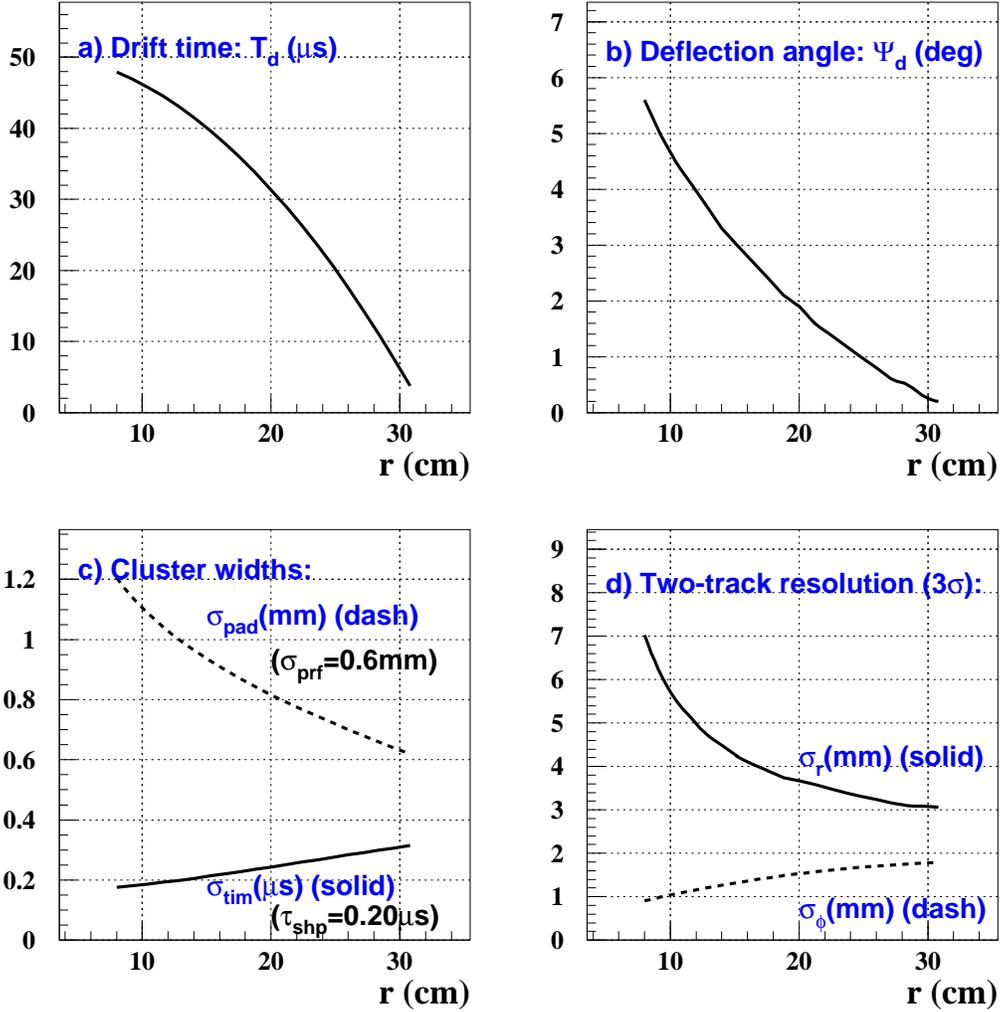,width=15.6cm}
\caption{Electron cluster properties as a function of the radial position
of the track hit in the FTPC. (a) drift time; (b) deflection angle;
(c) cluster widths in time ($\mu$s, solid line) and pad (mm, dashed line)
directions; (d) two-track resolution (3$\sigma$) in the radial (solid line)
and azimuthal (dashed line) directions.}
\label{fig2}
\end{figure*}

Due to the short drift length of only 23 cm a cool gas mixture with CO$_2$ or DME
can be used. It has a low diffusion coefficient for electrons and a small Lorentz
angle \cite{Bit97}. After extensive measurements an Ar/CO$_2$ ($50 \% / 50 \%$) mixture 
was selected which is nonflammable, shows no or little ageing effect in
comparison to hydrocarbons and is chemically less agressive  than a mixture with DME.
Fig. \ref{fig2} shows the measurements of drift time, cluster sizes, 
deflection angle due to the Lorentz force in the magnetic field 
and two track resolution for the Ar/CO$_2$ gas mixture \cite{Bie98}.

\subsection{Readout Chambers}

\begin{figure}
\epsfig{file=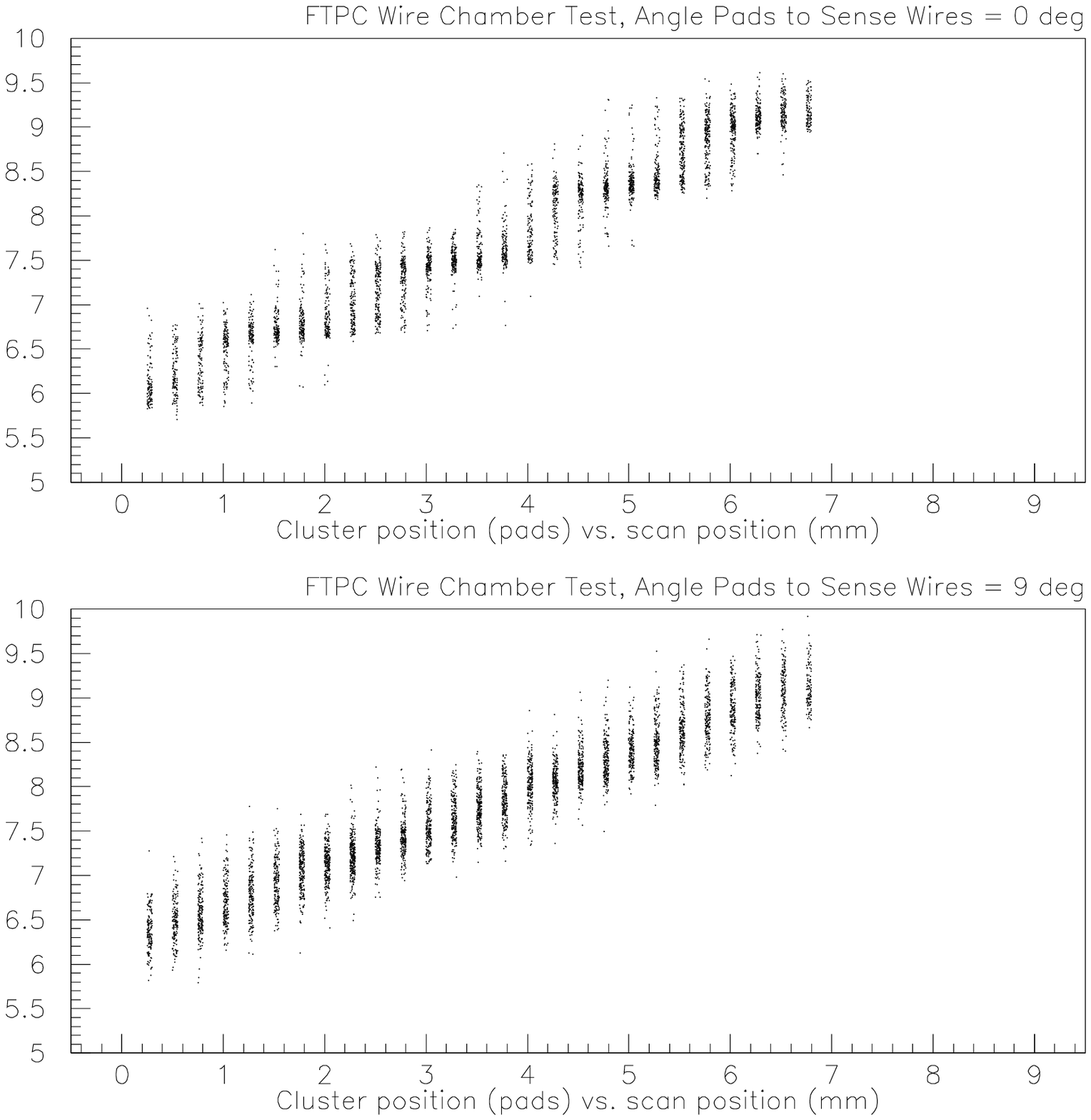,width=\linewidth}
\caption{Measured position of a laser beam for two crossing angles between
the anode wires and the pad axis. For $0^o$ (top) systematic shifts due
to the wire structure are observed, which disappear for $9^o$ angle (bottom).}
\label{fig3}
\end{figure}

\begin{figure*}
\begin{center}
\epsfig{file=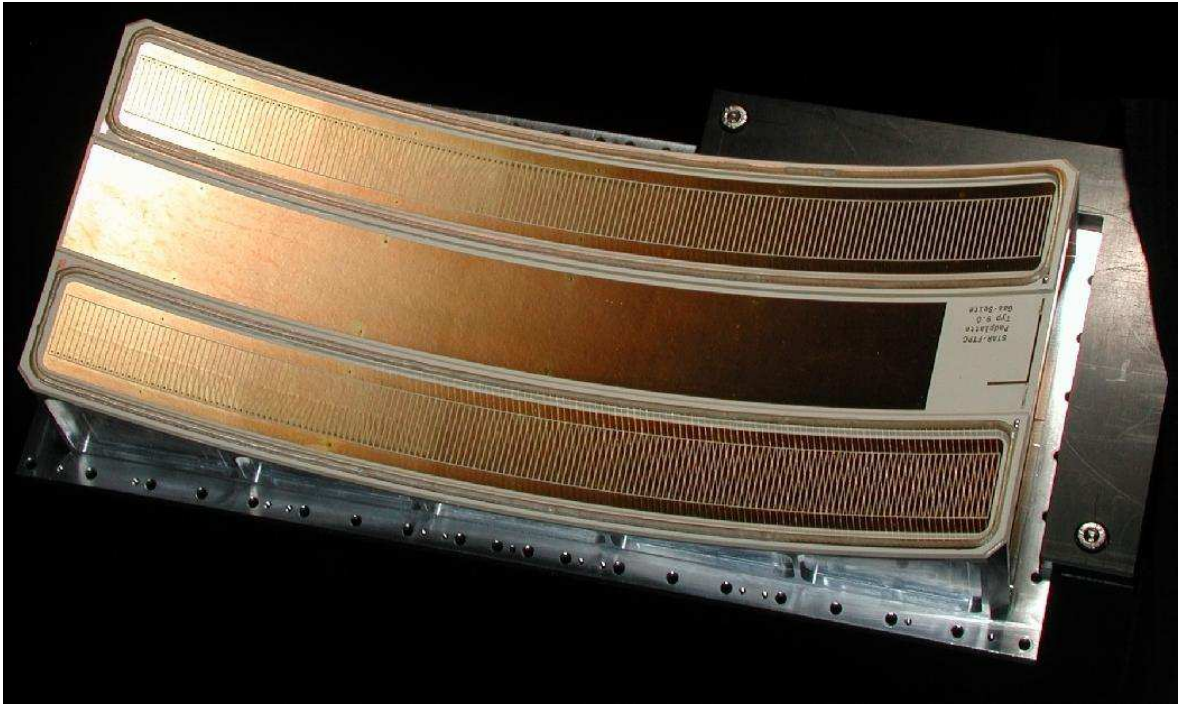,width=15.6cm}
\caption{Photograph of a FTPC readout chamber. The bending radius is 305 mm
and each of the two padrows has 160 pads.}
\label{fig4}
\end{center}
\end{figure*}

\begin{figure}
\epsfig{file=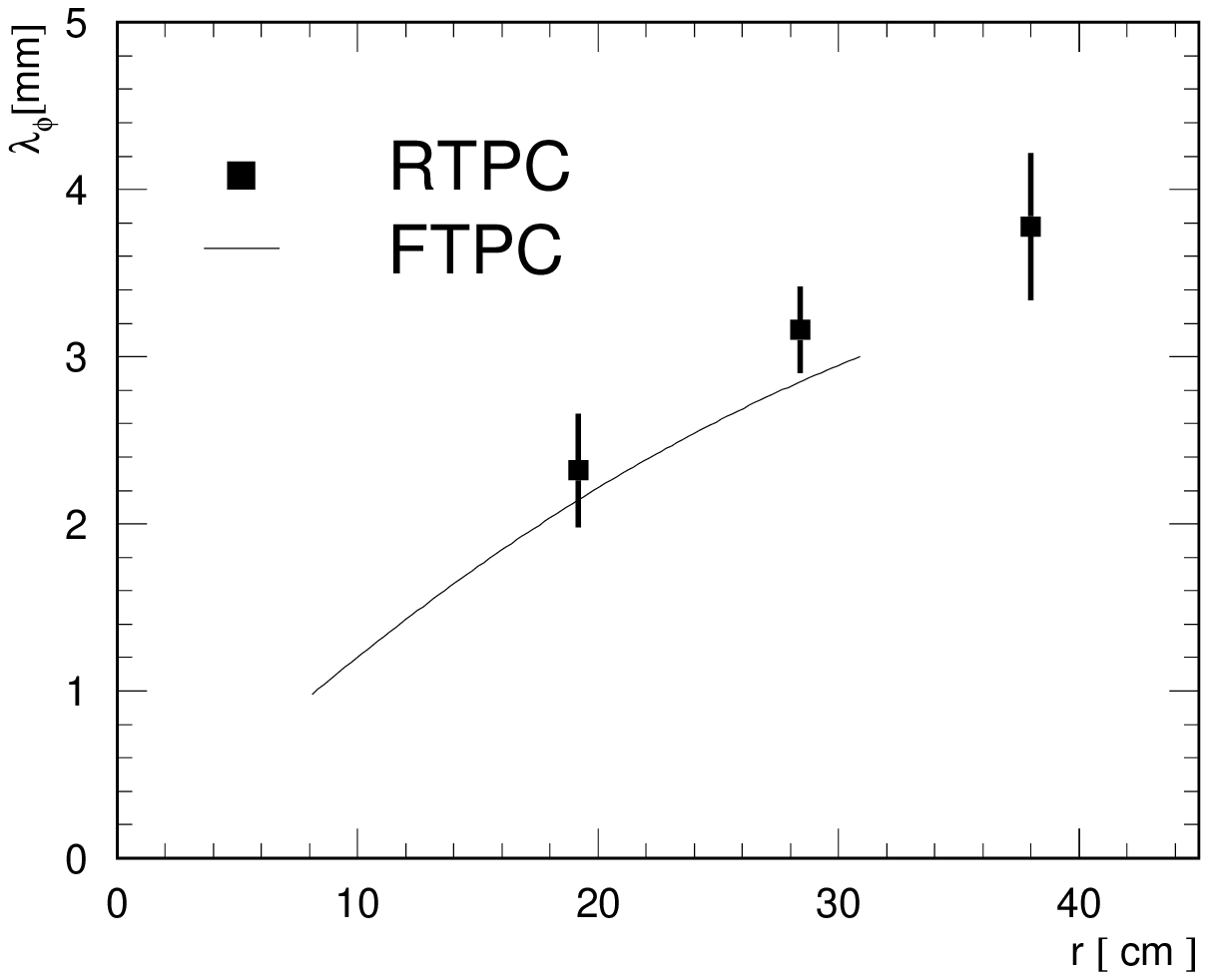,width=\linewidth}
\caption{Measured 2-track resolution in the prototype RTPC (data points) and
expected reolution in the FTPC (line) as a function of the radial distance
$r$ from the beam axis.}
\label{fig5}
\end{figure}

In a conventional TPC the anode (amplification) wires are orthogonal to the axial
direction of the pads. This is impossible in the case of a curved readout
chamber. The wires can not be parallel to the pads and therefore to the cylinder axis
either, because focusing effects then lead to periodic shifts in the position measurement.
This is demonstrated in fig. \ref{fig3}. However, if two or more wires
cross the pad under a small angle this effect already vanishes. 
For the FTPC design an angle of 17.4$^0$  was chosen resulting in three
wires crossing each pad for the selected pad-wire geometry. The anode wires are first glued
on the flat pad plate with conductive epoxy. Afterwards the plate is bent between
3 rollers to the final curvature without breaking the wires. A complete
readout chamber with 2 padrows is shown in fig. \ref{fig4}. With only
1.5 mm distance between the anode wires and the pad plane the spread of the signal
(the so-called Pad Response Function)  is of similar narrow width. This together 
with the low electron diffusion and the radial drift principle results in the required 
2-track separation of about 1 mm as can be seen in fig. \ref{fig5}. 

\subsection{Readout Electronics}

The two FTPCs have 19,200 channels of electronics, capable of measuring
the charge drifting to the readout chambers in short time samples. 
The drift time of about 50 $\mu$s for the 23 cm maximum drift length is
subdivided into 256 time bins. Because of the slow drift gas 
and the resulting long duration of the collection of the electron
cloud from a track crossing a shaping time of 350 ns
is used. The sampling rate is 5 MHz. The design of the front end 
electronics closely follows that of the central TPC \cite{Kle96}. Each pad is read
out by a low-noise STAR preamplifier/shaper (SAS), which sends signals to a
switched capacitor array/ADC chip (SCA/ADC). Four of these chip sets,
handling 64 channels, are mounted on a small FEE card, which is positioned
directly on the detector, parallel to the readout chambers. Fifteen FEE
cards are read out by a readout board, which sends the signals via a 1.2
Gbit/s fiber-optic link to the data acquisition system. The readout board
also controls the FEE cards, utilising signals from the clock and trigger
distribution system, and the slow control links. For maintaining the proper
operating temperature the
FEE and readout boards are water cooled using a leakless, low pressure
circulation system \cite{Bost}.
The FTPCs are remotely operated through a VME based supervision system.
Data logging and visualisation are performed by a software package
developed within the EPICS mainframe \cite{Kozu}.

\subsection{Laser Calibration System}

A laser calibration system serves the following primary purposes: 
\begin{itemize}
 \item Provides straight ionized tracks of known position to infer
corrections for spatial distortions caused by mechanical or drift field
imperfections. 
 \item Helps to calibrate the drift
       velocity in the non-uniform radial drift field.
 \item Tests the detector independent of collider operation.
\end{itemize}
The design of this system, similar to the STAR-TPC laser system
\cite{Aly96}, uses a frequency quadrupled Nd:YAG laser to provide a UV (266
nm) beam. The beam is expanded to 30 mm, split and transported via mirrors
to the two FTPC detectors. Remote angle control of two mirrors in each path
plus a CCD readout maintains precise steering of the two beams to the
detectors. At the detector the beam is subdivided with pickoff mirrors into
3, 8 mm beams which pass into the gas volume through fused silica
windows. Each of these beams are further split with smaller pickoff mirrors
into 5, 1 mm beams, producing a total of 15 fiducial, ionizing beams
distributed in the active volume.

\section{Simulation and Reconstruction of  Experimental Data}

The first step in the reconstruction of tracks is to calculate the track points
(cluster finding) from the charge distribution measured by the readout electronics. 
In a second step (track finding), these track points are grouped to tracks. Using
the magnetic field map, the up to ten position measurements per track are then
used to fit the momentum.

\subsection{Cluster Finding}

The reconstruction of track points is done by the FTPC cluster finding
program \cite{HUE01}. It is 
optimized to deal with high track densities while minimizing the use of computing
time. The program reads in the electronic signal data from the data acquisition
system, looks for areas of nonzero charge (cluster), deconvolutes clusters and
fits the point coordinates. The transformation from pad position and drift time
into cartesian coordinates includes the correction of distortions introduced by the
magnetic field. For a typical central Au+Au collision with 1000 particles in 
both FTPCs and an occupancy of $25 \%$ in the inner region the program
needs about 2 seconds on a 930 MHz Intel PentiumII processor.

\subsection{Track Reconstruction}

\begin{figure}
\epsfig{file=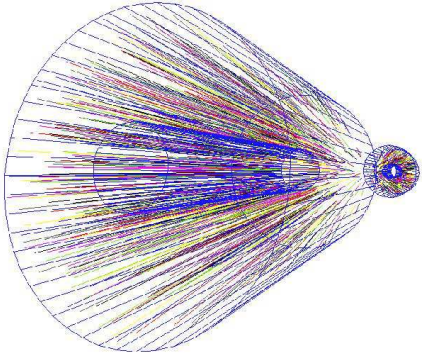,angle=90,width=\linewidth}
\caption{Reconstructed tracks in the FTPCs from a Au+Au collision at 
$\sqrt{s_{NN}} = 200 $ GeV simulated with the HIJING model. From 14,745
space points 1,026 tracks were reconstructed.}
\label{fig6}
\end{figure}

The second step in the analysis of FTPC data is the reconstruction of the particle
tracks and their momenta. The FTPC track reconstruction code is based on
an algorithm developed for fast online reconstruction \cite{JEP96}. It is a conventional 
track-following algorithm optimized for minimum use of computing power. In this code
all position calculations are done in a transformed coordinate system in which points
appear on a straight line if they form a helix in cartesian coordinates. This
processing step is known as conformal mapping \cite{JEP96}. It saves calculation time in the
track fitting, because all fits can be done by linear regression. After the track finding
step the code determines a primary vertex position by extrapolating and
intersecting all the reconstructed tracks.
Finally the particle momenta are fitted using the magnetic field map and the
vertex position. Fig. \ref{fig6} shows a reconstructed HIJING event for a central
Au+Au collision at 200 GeV per nucleon pair. From 14,745 space points 1,026
particle tracks were reconstructed in less than 2 seconds.

\section{Physics Simulation Studies}

\begin{figure}
\epsfig{file=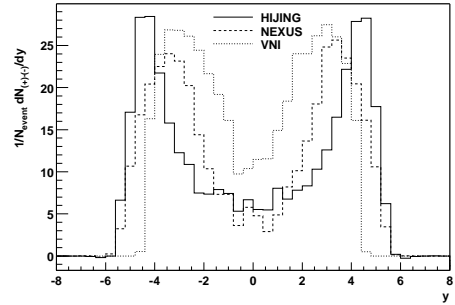,width=\linewidth}
\caption{Rapidity distribution of net positive charges (excess of
positively over negatively charged particles) in Au+Au collisions at
$\sqrt{s_{NN}}$ = 200 GeV simulated with three different models.}
\label{fig7}
\end{figure}

\begin{figure}
\epsfig{file=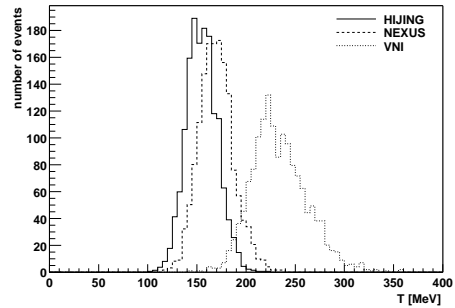,width=\linewidth}
\caption{Distributions of the temperature parameter $T$  reconstructed 
event-by-event in the FTPCs for three different event generators.}
\label{fig8}
\end{figure}

Simulation studies demonstrate the capability of distinguishing different
theoretical models of nucleus-nucleus collisions such as HIJING, NEXUS and VNI
with measurements in the FTPCs. Fig. \ref{fig7} shows the rapidity distribution of
net positive charges which follow the p - $\bar{\mathrm{p}}$ distribution and characterise
the baryon stopping in the reaction. Fig. \ref{fig8} shows histograms of the effective
temperature as determined event-by-event. Such measurements will be used to study and
search for fluctuations of event properties and to select special event classes.

\begin{figure*}
\epsfig{file=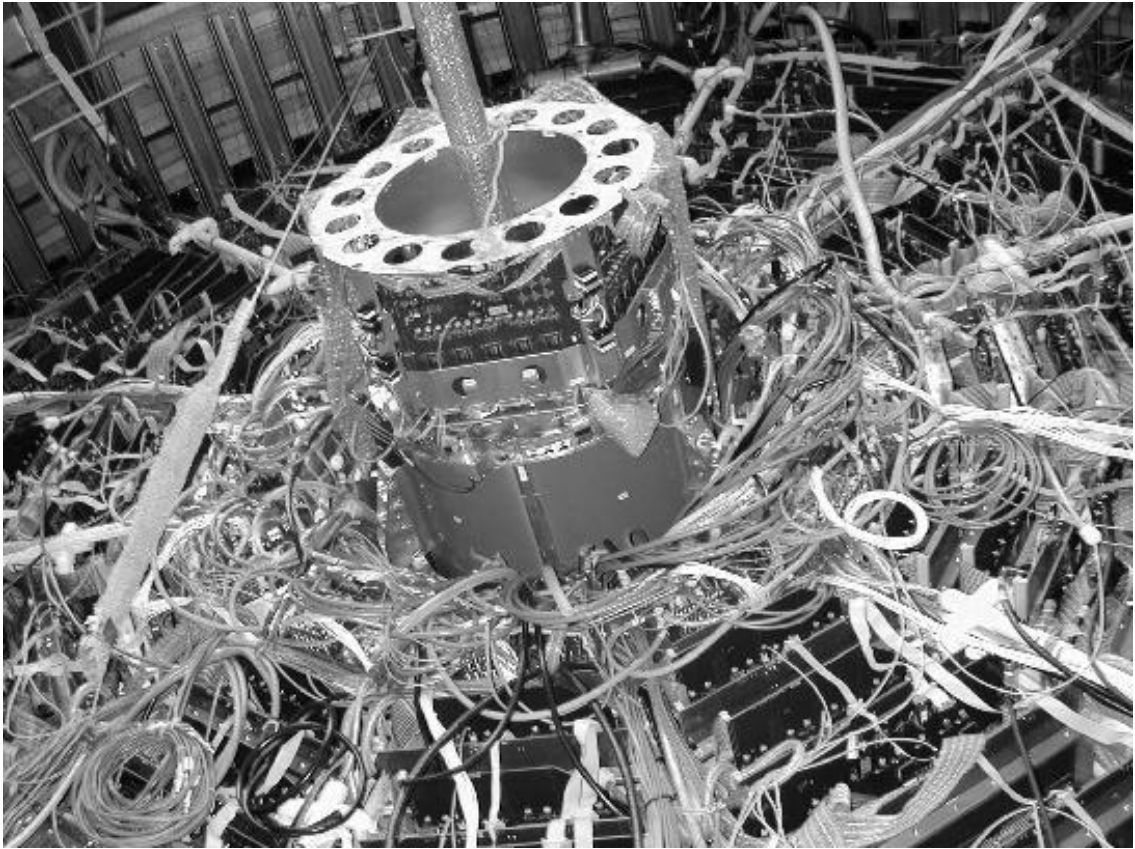,width=15cm}
\caption{One FTPC installed inside the STAR detector}
\label{fig9}
\end{figure*}

\section{Summary}

Based on the prototype measurements and simulations one expects to obtain a
position resolution of 100 $\mu$m, a two-track-separation of 1mm, a momentum
resolution between 12 and 15 $\%$, and an overall reconstruction efficiency between
70 and 80 $\%$. 

A summary of the main characteristics of the FTPCs is given in table \ref{par}.
The completed and tested FTPCs have been installed in the STAR experiment for
first data taking during summer and fall of 2001. One of them can be seen
mounted inside the STAR detector in Fig. \ref{fig9}.


\begin{center}
\begin{table*}
\caption{FTPC parameters} 
\begin{tabular}{||c||c||}
\hline \hline  
{\it\bf PARAMETER }& {\it\bf VALUE} \\ \hline \hline
\hline
{\bf Configuration}  &  { }   \\ \hline \hline
\# of TPC & 2 \\  \hline 
rows per TPC & 10  \\  \hline 
sectors per pad row & 6  \\  \hline 
pads per sector & $2 \times 160 $ \\  \hline 

\hline \hline
{\bf Sensitive Volume}  & {} \\   \hline\hline
inner radius & 8.0 cm \\   \hline
outer radius & 30.5 cm \\   \hline
chamber length & 120.0 cm ($150 < \left|z\right| < 270$ cm) \\   \hline
acceptance &  $2.5 < |\eta | < 4.0 (2.0^o < \theta < 9.3^o)$ \\   \hline

\hline \hline
{\bf Field Cage}  & {} \\   \hline\hline
drift cathode voltage & 10-15 kV \\   \hline
drift electrical field & 240-1400 V/cm (radial) \\   \hline
solenoid magnetic field &  0.5 T \\   \hline

\hline \hline
{\bf Gas}  & {}  \\   \hline\hline
gas mixture   & Ar(50\%)-CO$_2$(50\%)  \\   \hline
drift velocity  & 0.3 - 2.0 cm/$\mu$s  \\ \hline
trans. diffusion $D_T$ & 100-130 $\mu$m/$\sqrt{cm}$ \\   \hline
long. diffusion $D_L$ & 100-130 $\mu$m/$\sqrt{\rm cm}$ \\  \hline
Lorentz angle &  4 deg. (at 0.5 T)\\    \hline
gas gain & $\sim$ 1-2$\times 10^3$  \\     \hline

\hline \hline
{\bf Readout}  & {} \\   \hline\hline
\# of pads & 19200   \\   \hline
time bins per pad &  256 \\   \hline
pad pitch &  1.9 mm  \\   \hline
pad length & 20 mm \\   \hline
anodewire--pad gap & 1.5 mm \\   \hline
shaping time (FWHM) & 350 ns \\   \hline
SCA time bin size & 218 ns  \\   \hline
ADC dynamic range & 10 bits  \\   \hline

\hline \hline
\end{tabular}
\label{par}
\end{table*}
\end{center}



\clearpage

\begin{thebibliography}{99}

\bibitem{Bie98} F. Bieser et al., The Forward Time Projection Chamber for
the STAR Detector, {\em MPI PhE/98-3}, 1998

\bibitem{Sch99} A. Sch\"uttauf et al., A Forward TPC for STAR,
{\em Nucl. Phys.} {\bf A 661} (1999) 677c

\bibitem{Wie97} H. Wieman et al., STAR TPC at RHIC,  {\em IEEE
Trans. Nucl. Sci.} {\bf 44} (1997) 671

\bibitem{Bit97} X. Bittl et al., Diffusion and Drift Studies of Ar-DME/CO$_2$/CH$_4$
Gas Mixtures for a radial TPC in the E$\times$B Field,
 {\em Nucl. Instr. Meth.} {\bf A 398} (1997) 249

\bibitem{Kle96} S. Klein et al., Front-End Electronics for the STAR TPC,   {\em IEEE
Trans. Nucl. Sci.} {\bf 43} (1996) 1768

\bibitem{Aly96} M. Alyushin et al., Laser Calibaration System for STAR TPC,
{\em IEEE Nucl. Sc. Symp. Conference Record} {\bf 96CH35974} (1996),\\
Proceedings of the 9th Vienna Conference on Intrumentation, {\em Nucl. Instr. Meth.}
in print

\bibitem{HUE01} H. H\"ummler, Doctorate Thesis, Technical University,
M\"unchen, Germany, 2000, {\em MPI PhE/2001-04} 2001,\\
H. H\"ummler, Simulation and Reconstruction of Data from the STAR FTPC,
{\em STAR Note SN0429} (2000)

\bibitem{JEP96} P. Yepes, A fast track pattern recognition, {\em
Nucl. Instr. Meth.} {\bf A 380} (1996) 582\\
M.Oldenburg, Doctorate Thesis, Technical University,
M\"unchen, Germany, 2001, in preparation

\bibitem{Bost} M. Bosteels, Cern internal report, CERN/LHCC/99-33 (1999)

\bibitem{Kozu} A.J. Kozubal et al., ICALEPS89 Proceedings, 288, (Vancouver, 1989)

\end{thebibliography}
\end{document}